\documentclass{sig-alternate}

\usepackage{graphicx}
\usepackage{url}
\usepackage{mdwlist} 
\usepackage{times}
\usepackage{dblfloatfix}

\def\onethird{0.25\textwidth}

\def\sharedaffiliation{
\end{tabular}
\begin{tabular}{c}}

\begin{document}

\title{Agents, Bookmarks and Clicks:\\
A topical model of Web navigation}

\numberofauthors{3}
\author{
\alignauthor Mark R. Meiss$^{1,3}$\titlenote{Corresponding author.
Email: \texttt{mmeiss@indiana.edu}}
\alignauthor Bruno Gon\c{c}alves$^{1,2,3}$
\alignauthor Jos\'{e} J. Ramasco$^{4}$
\and
\alignauthor Alessandro Flammini$^{1,2}$
\alignauthor Filippo Menczer$^{1,2,3,4}$
\sharedaffiliation
\affaddr{$^1$School of Informatics and Computing, Indiana University, Bloomington, IN, USA}\\
\affaddr{$^2$Center for Complex Networks and Systems Research, Indiana University, Bloomington, IN, USA}\\
\affaddr{$^3$Pervasive Technology Institute, Indiana University, Bloomington, IN, USA}\\
\affaddr{$^4$Complex Networks and Systems Lagrange Laboratory (CNLL), ISI Foundation, Turin, Italy}
}

\maketitle



\begin{abstract} 
Analysis of aggregate and individual Web traffic has shown that PageRank 
is a poor model of how people navigate the Web.  Using
the empirical traffic patterns generated by a thousand users, 
we characterize several properties of Web traffic that cannot be reproduced 
by Markovian models. We examine both aggregate statistics capturing collective 
behavior, such as page and link traffic, and individual statistics, 
such as entropy and session size.  No model currently explains all of these 
empirical observations simultaneously.  We show that all of these traffic 
patterns can be explained by an agent-based model that takes into 
account several realistic browsing behaviors.  First, agents maintain 
individual lists of bookmarks (a non-Markovian memory mechanism) that are used as 
teleportation targets.  Second, agents can retreat along visited links, a branching mechanism that
also allows us to reproduce behaviors such as the use of a back button and tabbed browsing.
Finally, agents are sustained by visiting novel pages of topical interest, with adjacent pages being more
topically related to each other than distant ones.  This modulates the probability that an agent continues to browse 
or starts a new session, allowing us to recreate heterogeneous session 
lengths.  The resulting model is capable of reproducing
the collective and individual behaviors we observe 
in the empirical data, reconciling the narrowly focused 
browsing patterns of individual users with the extreme heterogeneity 
of aggregate traffic measurements.  This result allows us to identify a 
few salient features that are necessary and sufficient to interpret the browsing patterns
observed in our data.  In addition to the descriptive and explanatory power of 
such a model, our results may lead the way to more sophisticated, realistic, 
and effective ranking and crawling algorithms. 
\end{abstract}

\pagebreak 





\section{Introduction}

Despite its simplicity, PageRank~\cite{Brin98} has been a remarkably 
robust model of human Web browsing characterizing it as a random surfing activity. 
Such models of Web surfing have allowed us to speculate how people interact with the
Web.  As ever more people spend a growing portion of their time online, their
Web traces provide an increasingly informative window into human
dynamics. 
The availability of large volumes of Web traffic data enables 
systematic testing of PageRank's underlying navigation
assumptions~\cite{meiss08}.  Traffic patterns aggregated across 
users have revealed that some of its key assumptions---uniform random walk
and uniform random teleportation---are widely violated, making
PageRank a poor predictor of traffic.  Such results leave open the 
question of how to design a better Web navigation model.
Here we expand on our previous empirical analysis~\cite{meiss08,meiss09} 
by considering also \emph{individual} traffic patterns~\cite{goncalves08}.
Our results provide further evidence for the limits of simple (memoryless) 
Markovian models such as PageRank.  They suggest the need for an
\emph{agent-based} model with more realistic features, such as memory and 
topicality, to account for both individual and aggregate traffic patterns 
observed in real-world data.

Models of user browsing also have important practical applications.  First, 
the traffic received by pages and Web sites has a direct impact on the financial
success of many companies and institutions.  Indirectly, understanding traffic 
patterns has consequences for predicting advertising revenues and on policies 
used to establish advertising prices~\cite{douglis07pr}. 
Second, realistic models of Web navigation could guide the behavior of 
intelligent crawling algorithms, improving the coverage of important sites 
by search engines~\cite{Cho98, Menczer02TOIT}.  Finally, improved traffic models 
may lead to enhanced search ranking algorithms~\cite{Brin98, radlinski07, Liu08browserank}.

\subsection*{Contributions and outline}


In the remainder of this paper, after some background on related and prior 
work, we describe a data set collected through a field study from 
over a thousand users on the main campus of Indiana University. 
We previously introduced a model of browsing behavior, 
called \emph{BookRank}~\cite{goncalves09}, which extends PageRank 
by adding a memory mechanism. Here we 
introduce a novel agent-based model, which also 
accounts for the topical interests of users. We 
compare the traffic patterns generated by these models 
with both \emph{aggregated} and 
\emph{individual} Web traffic data from our field study.   
Our main contributions are summarized below:

\begin{itemize}


\item
We show that the empirical diversity of pages visited by individual users, 
as measured by Shannon entropy, is not well 
predicted by either PageRank or BookRank. 
This suggests that a typical user has both 
focused interests and recurrent habits, meaning that the diversity apparent
in many aggregate measures of traffic must be a consequence of the 
diversity across individual interests.

\item
When we build \emph{logical sessions} by assembling requests based on referrer information and
initiating sessions based on link-independent jumps~\cite{meiss09}, we find that  
endowing the model with a simple memory mechanism such as bookmarks (as in the Book\-Rank 
model) is sufficient to correct the mismatch between PageRank and the distributions 
of aggregate measures of traffic, but not to capture the broad distributions of 
individual session size and depth.

\item
We introduce an agent-based navigation model, \emph{ABC}, 
with three key, realistic ingredients: (1) \emph{bookmarks} are managed and used 
as teleportation targets, defining boundaries between logical sessions 
and allowing us to capture the diverse popularity of starting pages; 
(2) a \emph{back button} is available to account for tabbed browsing and explain 
the branching observed in empirical traffic; and (3) \emph{topical interests} 
drive an agent's decision to continue browsing or start a new session, 
leading to diverse session sizes.  The model also takes into consideration 
the \emph{topical locality} of the Web, so that an interesting page is likely to 
link to other interesting pages.


\item
Finally, we demonstrate that the novel ingredients of ABC allow
it to match or exceed PageRank and BookRank in reproducing the empirical
distributions of page traffic, link traffic, and of the popularity of session starting pages, 
while outperforming both PageRank and BookRank in
modeling user traffic entropy  
and size and depth of sessions.

\end{itemize}

\section{Background}

There have been many empirical studies of Web traffic patterns.  The
most common approach is the analysis of Web server logs.  These have
ranged from small samples of users from a few arbitrarily
selected Web server logs~\cite{catledge95} to large samples of users from
the logs of large organizations~\cite{goncalves08}.  One advantage
of this methodology is that it allows us to distinguish individual
users though their IP addresses (even if they may be anonymized), thus
capturing \emph{individual} traffic patterns~\cite{goncalves08}.
Conversely, the methodology has the drawback of biasing both the
sample of users and the sample of the Web graph being observed based
on the choice of target server.

An alternative source of Web traffic data is browser toolbars, which
gather traffic information based on the surfing activity of many
users.  While the population is larger in this scenario, it is still
biased by users who have opted to install a particular piece of
software.  Moreover, traffic information from toolbars is not
generally available to researchers.  Adar \textit{et
al.}~\cite{Adar08} used this approach to study the patterns of
revisitation to pages, but did not consider whether pages are
revisited within the same session or across different sessions.
A related approach is to identify a panel of users based on
desired characteristics, then ask them to install tracking
software.  This eliminates many sources of bias but incurs significant
experimental costs.  Such an approach has been used to describe the
exploratory behavior of Web surfers~\cite{beauvisageHT09}.
These studies did not propose models to explain the observed traffic patterns.

The methodology adopted in the study reported here captures traffic
data directly from a running network.  This approach was first adopted by Qiu
\textit{et al.}~\cite{Cho05WebDB}, who used captured HTTP packet
traces in the UCLA computer science department to investigate how
browsing behavior is driven by search engines.  Our study relies on a
larger sample of users.

One of the important traffic features we study here is the statistical
characterization of browsing sessions.  A common assumption is that
long pauses correspond to breaks between sessions.  Based on this
assumption, many researchers have relied on timeouts as a way of
defining sessions, a technique we have recently found to be
flawed~\cite{meiss09}.  This has led to the definition of
time-independent \emph{logical sessions,} based on the reconstruction
of session trees rooted at pages requested without a referrer.  The
model presented here is in part aimed at explaining the broad
distributions of size and depth empirically observed for these logical
sessions.

An aspect of Web traffic that has not received much attention in the
literature is the role of page content in driving users' browsing
patterns.  A notable exception is a study of the correlation between
changes in page content and revisit patterns~\cite{Adar09}.

On the modeling side, the most basic notion of Web navigation is that
users move erratically, performing a random walk through pages in the
Web graph.  PageRank~\cite{Brin98} is a random walk modified by the
process of teleportation (random jumps), modeling how users start new
browsing sessions by a Poissonian process with uniformly random
starting points.  This Markovian process has no memory, no way to
backtrack, and no notion of user interests or page content.  The
stationary distribution of visitation frequency generated by PageRank
can be compared with empirical traffic data.  We have shown that the
fundamental assumptions underlying Page\-Rank---uniform link selection,
uniform teleportation sources and targets---are all violated by actual
user behavior, making Page\-Rank a poor model of actual
users~\cite{meiss08}.  Such results leave open the question of how to
design a better Web navigation model.  That is the goal of the present
paper; we use such a random walk as a null model to evaluate our
alternative model.

More realistic models have been introduced in recent years to capture
potentially relevant features of real Web browsing behavior, such as
the back button~\cite{mathieu04,bouklit05}.  There have also been
attempts to model the role of the interplay between user interests and
page content in shaping browsing patterns.  Huberman \textit{et al.}
proposed a model in which pages visited by a user have interest values
described by a random walk; the navigation continues as long as the
current page has a value above a threshold~\cite{Huberman98}.  This
kind of model is closely related to algorithms designed to improve
topical crawlers~\cite{Menczer97b, Menczer00, Menczer02TOIT}.

We previously proposed a model in which the users maintain a list of 
bookmarks from which they start new sessions, providing memory of 
previously visited pages~\cite{goncalves09}.
We called this model \emph{BookRank,} since the bookmark selection is
carried out according to a ranking based on the frequency of visits to
each bookmark.  
This model is able to reproduce a fair number of
characteristics observed in empirical traffic data, including the page
and link traffic distributions.  Unfortunately, BookRank fails to
account for features related to the navigation patterns of individual
users, such as entropy and session characteristics.  This failure is 
not remedied by the introduction of a back button into the model. In the
remainder of this paper, we extend the BookRank model to address these
shortcomings.

\section{Empirical traffic data} 
\label{study}

\subsection{Data acquisition}

The HTTP request data we use in this study was gathered from a
dedicated FreeBSD server located in the central routing facility of
the Bloomington campus of Indiana University~\cite{meiss09}.
This system had a 1~Gbps Ethernet
port that received a mirror of \emph{all} outbound network traffic from one
of the undergraduate dormitories.  This dormitory 
is home to just over a thousand
undergraduates 
split roughly evenly between men
and women. 
To the best of our knowledge this is the largest population sample 
whose \emph{every} click has been recorded and studied over an 
extended period of time.


To identify individual requests, we first capture only packets
destined for TCP port 80.  While this does eliminate Web traffic
running on non-standard ports, it allows for an improved rate of
capture that more than offsets the lost data.  We make no attempt to
capture or analyze encrypted (HTTPS) traffic using TCP port 443.  For
each packet we capture, we use a regular expression search to
determine whether it contains an HTTP GET request.  If we do find a
request, we analyze the packet further and
log the request, making note of the MAC address of the client,
a timestamp, the virtual host and path requested, the referring URL,
and a flag indicating whether the user agent matches a mainstream
browser.  We record the MAC address only in order to distinguish the
traffic of individual users.  We thus assume that most computers have
a single primary user, which is reasonable: most students own 
computers, and only a few public workstations are available
in the dormitory.  Furthermore, as long as users do not replace their
network interface card, this information remains constant.

The aggregate traffic was low enough to permit full rate of collection
without dropping packets.  While this collection system offers a rare
opportunity to capture the complete browsing activity of a large
population, we do recognize some potential disadvantages.  Because we
do not perform TCP stream reassembly, we can only analyze requests
that fit in a single Ethernet frame.  The vast majority of requests do
so, but some GET-based Web services do generate extremely long URLs.
Without stream reassembly, we also cannot log the Web server's
response to each request, making us unaware of failed requests and
redirects.  A user can spoof the HTTP referrer field; we assume that
few students do so.  Finally, although they are in a residential
setting, the students are at an academic institution and represent a
biased sample of the population of Web users at large.  This is an
inevitable consequence of any local study of a global and diverse
system such as the Web.

The click data was collected over a period of about two months, from
March~5, 2008 through May~3, 2008.  This period included a
week-long vacation during which no students were present in the
building.  During the full data collection period, we detected nearly
408 million HTTP requests from a total of 1,083 unique MAC
addresses.

Only a minority of HTTP requests reflect an actual human being trying
to fetch a Web page for display.  We retain only requests 
that are likely to be 
for actual Web pages, as opposed to
media files, style sheets, Javascript code, images, and so forth.  We
make this determination based on the extension of the URL requested,
which is imprecise but a reasonable heuristic in the absence of access
to the MIME type of the server response.  We also filtered out a small
subset of users with negligible (mostly automated) activity. 
Finally, we removed 
some spoofed requests generated by an 
anonymization service that attempted to obscure traffic to an adult
chat site.


%
%

Privacy concerns and our agreement with the Human Subjects Committee
of our institution also obliged us to 
strip off all identifiable query parameters from the
URLs.  Applying this anonymization procedure affects roughly one-third
of the remaining requests.  This procedure means that two URLs with
different CGI variables will be treated as the same.  While this is a
mistaken assumption for sites in which the identity of the page being
requested is a query parameter, it helps in the common case that the
parameters affect some content within a largely static framework.

%

\begin{table}
\caption{Approximate dimensions of the filtered and anonymized data set.}
\begin{center}
\begin{tabular}{lr}
\hline
Page requests & 29,494,409 \\
Unique users & 967 \\
Unique URLs & 2,503,002 \\
Unique target URLs & 2,084,031 \\
Unique source URLs & 864,420 \\
Number of sessions & 11,174,254 \\
Mean sessions per user & 11,556 \\
\hline
\end{tabular}
\end{center}
\label{table:data}
\end{table}

Once we have a filtered set of HTTP requests (``clicks''), we organize
each user's clicks into a set of sessions.  These sessions are not
based on a simple timeout threshold; our prior work demonstrates that
most statistics of timeout-based sessions are functions of the
particular timeout used, which turns out to be
arbitrary~\cite{meiss09}.  Instead, we organize the clicks into
tree-based \textit{logical} sessions using the referrer information
associated with each request, according to an algorithm described formally in
our previous work~\cite{meiss09}.  The basic notions are  
that new sessions are initiated by requests with an empty referrer field; 
that each request represents a directed edge from a referring URL to a target
URL; and that requests are assigned to the session in which their
referring URL was most recently requested.

The session trees built in this way offer several advantages.  First,
they mimic the multitasking behavior of users in the age of tabbed
browsing: a user may have several active sessions at a time.  Second,
the key properties of these session trees, such as size and depth, are
relatively insensitive to an additional timeout constraint introduced
for the sake of plausibility~\cite{meiss09}.  In the current analysis, we impose a
half-hour timeout as we form the sessions: a click cannot be
associated with a session tree that has not received additional
requests within thirty minutes.

Most importantly, the tree structure allows us to infer how users
backtrack as they browse.  Because modern browsers follow
sophisticated caching mechanisms to improve performance, unless
overridden by HTTP options, a browser will generally not issue another
request for a recently accessed page.  This prevents us from observing
multiple links pointing to the same page (within a single logical
session) and gives us no \textit{direct} way of determining when the
user presses the back button.  However, session trees allow us to
\textit{infer} information about backwards traffic: if the next
request in the tree comes from a URL other than the most recently
visited one, the user must have navigated to that page, or opened it in a separate tab.

The dimensions of the resulting data set are shown in
Table~\ref{table:data}.  In \S~\ref{datafeatures}, we present the most
relevant properties of this data for the discussion that follows; more
detailed analysis of the empirical sessions can be found
in~\cite{meiss09}.

\subsection{Data descriptors}
\label{datafeatures}

Any statistical description strives to achieve a compromise between
the need to summarize the behavior of the data and the need to
describe such behavior accurately.  In the case of many human
activities, including those on the Web, we know that the data does not
behave in a normal (Gaussian) fashion, but rather fits into
heavy-tailed distributions approximated best by power
laws~\cite{Broder00, meiss08}.  In many cases, the mean
and median are not a sufficient description of the data, as shown by a
large and diverging variance and heavy skew.  The next best
description of any quantity is a histogram of its values.  We
therefore present these distributions in terms of their estimated
probability density functions rather than measures of central
tendency. To characterize the properties of our traffic data and 
evaluate the models proposed later in this paper, 
we focus on the distributions of the six quantities outlined below.

\begin{description}

\item[Page traffic]
The total number of visits to each page.  Because of caching mechanisms, 
the majority of revisits to a page by a single user beyond the first 
visit within each session will not be represented in the data.

\item[Link traffic]
The total number of times each link between pages has been traversed
by a user, as identified by the referrer and destination URLs in each
request.  Again, because of caching behavior, we typically observe 
only the first click to a destination page within each session.

\item[Empty referrer traffic]
The number of times each page is used to initiate a new session.  We
assume that a request without a referring page corresponds to the user
initiating a new session by using a bookmark, opening a link from
another application, or manually entering an address.

\item[Entropy]
Shannon information entropy.  For an individual user $j$, the entropy
is defined as $S_j = -\sum_i \rho_{ij} \log_2 \rho_{ij}$ where
$\rho_{ij}$ is the fraction of visits of user $j$ to site $i$ aggregated across sessions.

\item[Session size]
The number of unique pages visited in a logical session tree.


\item[Session depth]
The maximum tree distance between the starting page of a session and
any page visited within the same session.  (Recall that session graphs
have a tree-like structure because requests that go back to a
previously visited page are usually served from the browser cache.)


\end{description}

We have already characterized some of these distributions in
preliminary work~\cite{goncalves09,meiss09}.  Another feature
sometimes used to characterize random browsing behavior is the
distribution of return time, which in this case would be the number of
clicks between two consecutive visits to the same page
by a given user~\cite{goncalves08,goncalves09}.  However, cache
behavior and overlapping sessions mean that this
information cannot be retrieved in a reliable way from the empirical
data.

\subsection{Reference models}
\label{ref-models}


\begin{figure}
\centerline{\includegraphics[width=\columnwidth]{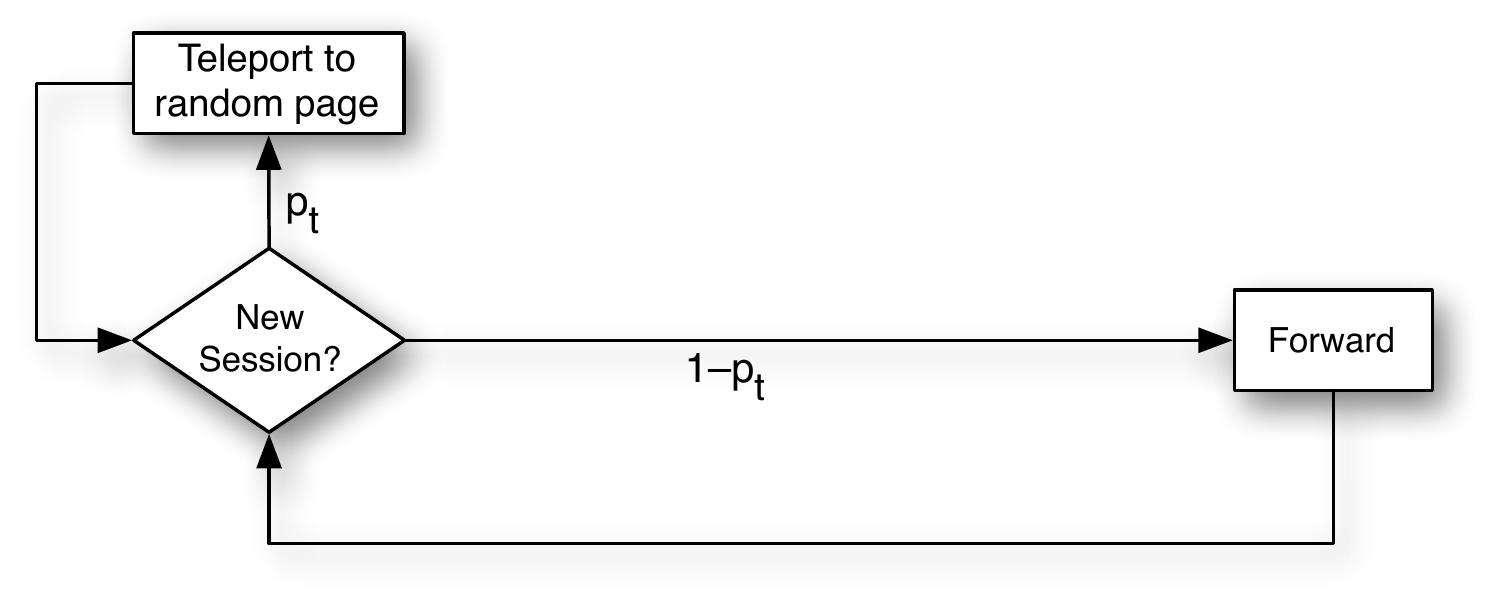}} 
\caption{Schematic illustration of the PageRank model.}
\label{PR-model}
\end{figure}

\begin{figure}
\centerline{\includegraphics[width=\columnwidth]{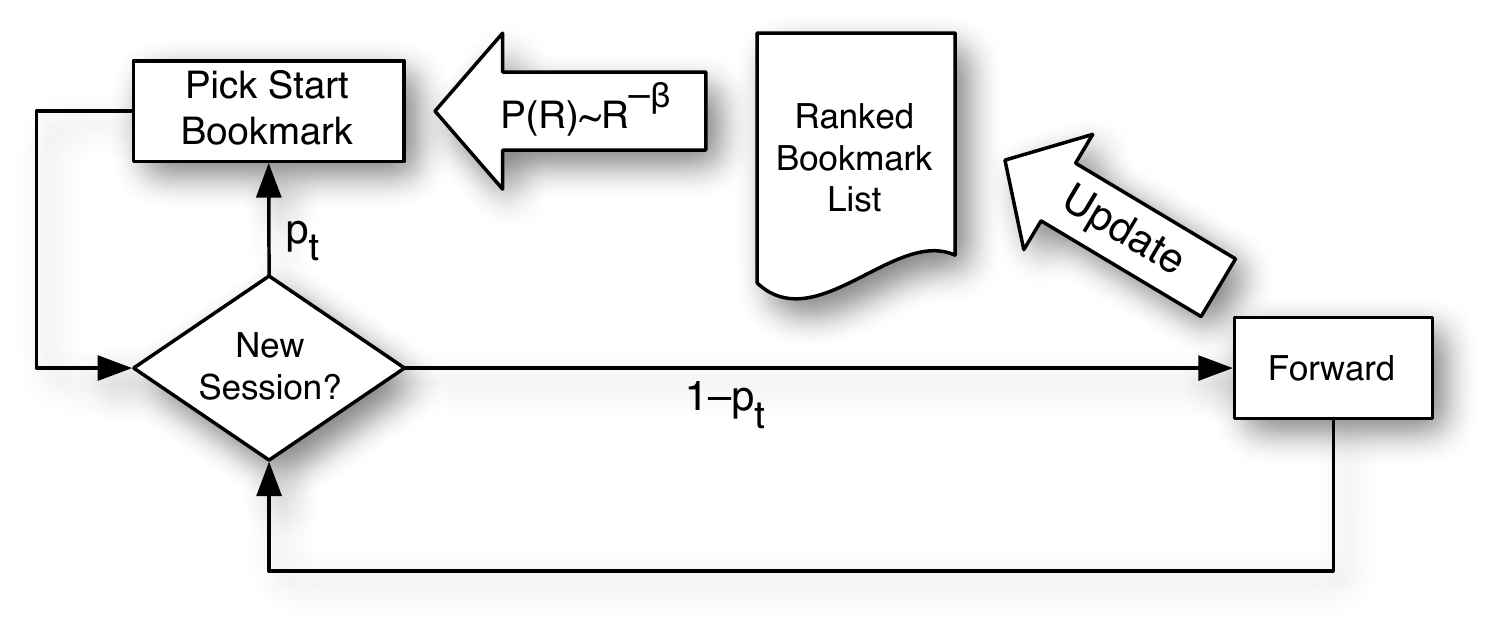}} 
\caption{Schematic illustration of the BookRank model.}
\label{BR-model}
\end{figure}

To properly analyze these distributions, we compare them with those
generated by two reference models based on PageRank-like modified
random walkers with teleportation probability $p_t = 0.15$.  To obtain
a useful reference model for traffic data that is based on
individuals, we imagine a population of PageRank random walkers, as
many as the users in our study. The first reference model (PageRank) 
is illustrated in Fig.~\ref{PR-model}. Each walker browses for as many sessions
as there were empirical sessions for the corresponding real-world
user.  The PageRank sessions are terminated by the
constant-probability jumps, so the total number of pages visited by a
walker may differ from the corresponding user. Teleportation jumps lead 
to session-starting pages selected uniformly at random. 

The second reference model (BookRank) is illustrated in Fig.~\ref{BR-model}. 
The key realistic ingredient that differentiates this model from Page\-Rank is 
related to memory: agents maintain individual lists
of bookmarks that are chosen as teleportation targets based on the
number of previous visits. 
Initially, each agent randomly selects a starting page (node).  
Then, agents navigate the Web by repeating the following steps: 

\begin{enumerate}

\item
With probability $1-p_{t}$, the agent navigates locally,
following a link from the present node selected with uniform probability. 
Unless previously visited, the new node is added to the bookmark list.
The frequency of visits is recorded, and the list of bookmarks is  
kept ranked from most to least visited.  

\item
Otherwise, with probability $p_{t}$, the agent teleports (jumps) 
to a previously visited page (bookmark).  The bookmark with rank $R$ 
is chosen with probability $P(R) \propto R^{-\beta}$. 

\end{enumerate}

The above mechanism mimics the use of frequency ranking in 
various features of modern browsers, such as URL completion in the address 
bar and suggested starting pages in new windows. 
The functional form $P(R)$ for the bookmark choice is motivated by data on selection  
among a ranked list of search results~\cite{Fortunato05egalitarian}. 

%
%

In our 
simulations, browsing occurs on scale-free networks with 
$N$ nodes and degree distribution $P(k) \sim k^{-\gamma}$, generated
according to the growth model of Fortunato \textit{et al.}~\cite{fortunato06}.  
We used a large graph with $N = 10^7$ nodes to ensure that the network 
would be larger than the number 
of pages visited in the empirical data (cf.~Table~\ref{table:data}). 
We also set $\gamma = 2.1$ to match our data set. 
This graph is constructed with symmetric links to prevent dangling 
links; as a result, each node's in-degree is equal to its out-degree.

Within a reference model's session, we simulate the browser's cache by
recording traffic for links and pages only when the target page has
not been previously visited in the same session.  This way we can
measure in the models the number of unique pages visited in a session,
which we can compare with the empirical session size.  We assume that
that cached pages are reset between sessions.

\subsection{Data analysis}
\label{analysis}

\begin{figure}
\centerline{\includegraphics[width=\columnwidth]{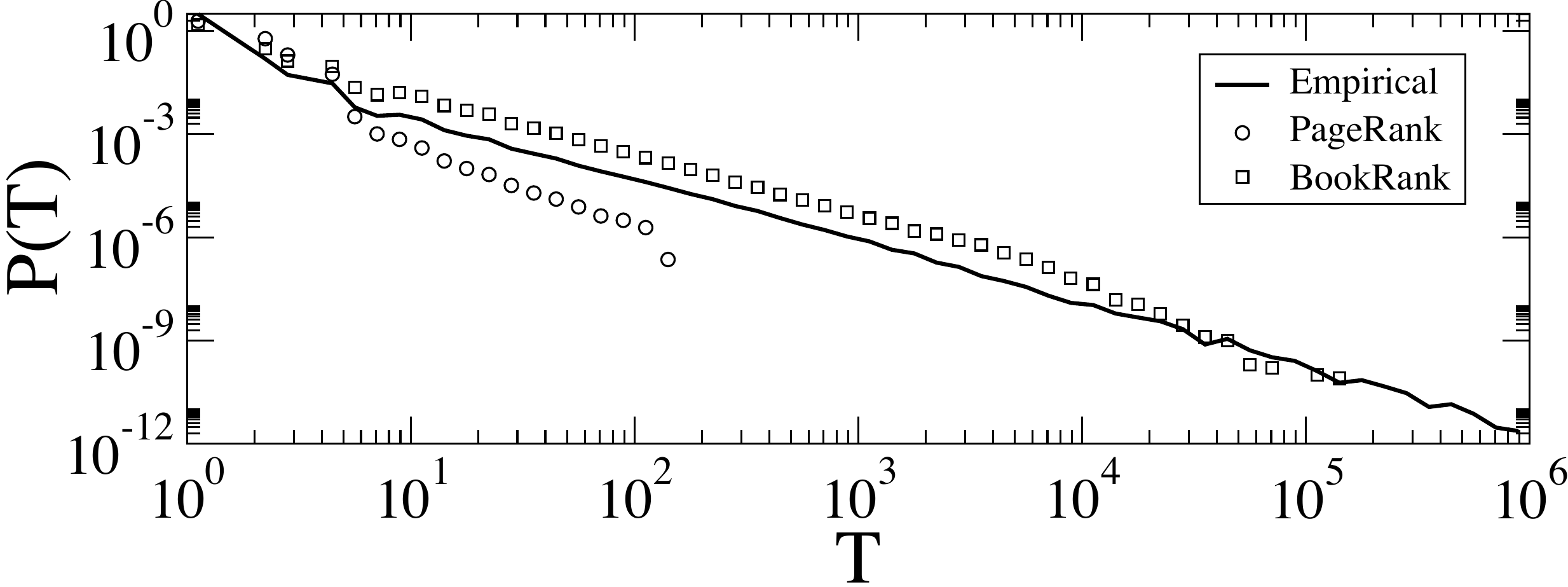}}
\caption{Empirical distribution of page traffic versus baselines.}
\label{fig:traffic}
\end{figure}

We first consider the aggregate distribution of traffic received by
individual pages, as shown in Fig.~\ref{fig:traffic}.  The empirical
data show a very broad power-law distribution for page traffic, 
$P(T) \sim T^{-\alpha}$, with exponent $\alpha \approx 1.75$,%
\footnote{The fact that $\alpha <2$ is significant: in this case,
both the variance and the mean of the distribution diverge in the
limit of an infinite-size network.}
which is consistent with our prior results for host-level
traffic~\cite{meiss08,meiss09}.

Theoretical arguments~\cite{noh04} suggest that PageRank should behave 
in a similar fashion. If we disregard 
teleportation, a node of in-degree $k$ may expect a visit if one
of its neighbors has been visited in the previous step.  The traffic it
will receive will be therefore proportional to its degree, if no
degree-degree correlations are present in the graph.  This intuition, 
as well as prior empirical results~\cite{meiss08}, lead us to expect 
that PageRank's prediction of the distribution of traffic received by
a Web page is described by a power law $P(T) \sim T^{-\alpha}$
where $\alpha \approx 2.1$ is the same exponent observed in the
distribution of the in-degree~\cite{Broder00,Fortunato05topten}. 
Indeed this is consistent with the distribution generated by the Page\-Rank 
reference model in Fig.~\ref{fig:traffic}. On the other hand, the 
traffic generated by BookRank is biased toward previously 
visited pages (bookmarks), and therefore has a broader distribution 
(by three orders of magnitude) in better agreement with the empirical 
data, as shown in Fig.~\ref{fig:traffic}. 

\begin{figure}
\centerline{\includegraphics[width=\columnwidth]{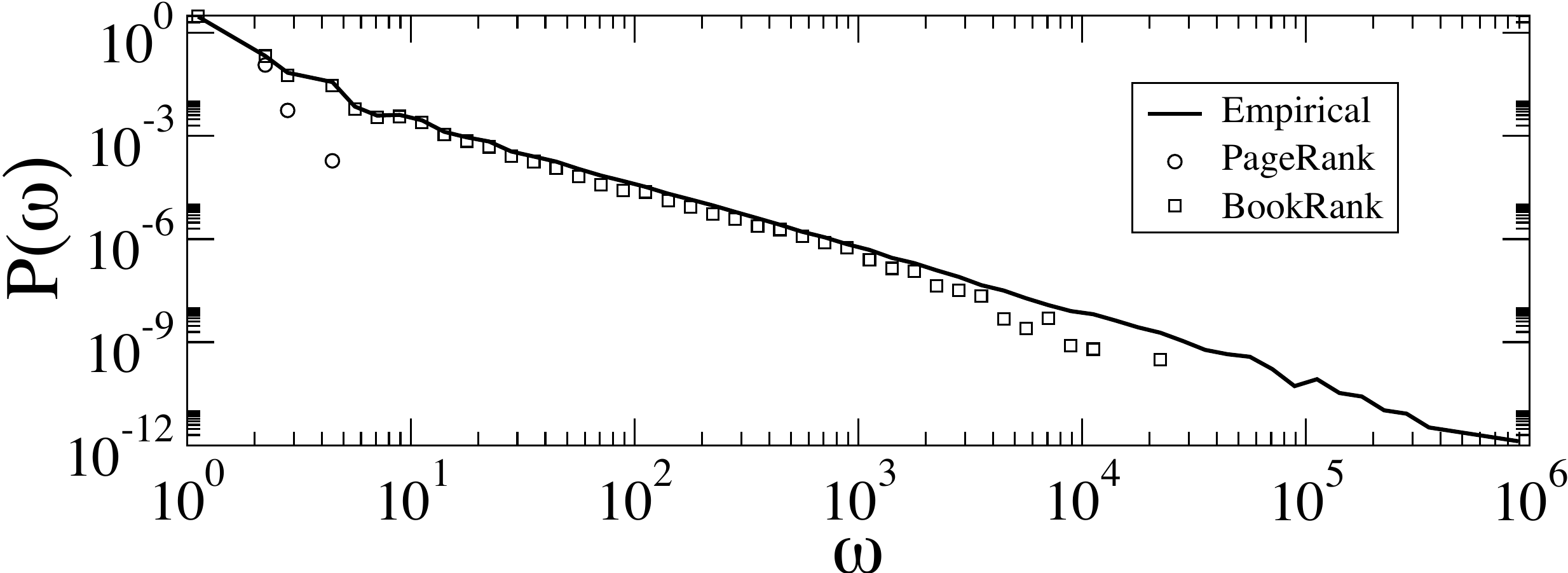}}
\caption{Empirical distribution of link traffic versus baselines.}
\label{fig:traf_link}
\end{figure}

The distribution of weights $\omega$ across links between pages allows
us to consider the diversity of traffic crossing each hyperlink in the
Web graph.  In Fig.~\ref{fig:traf_link}, we compare the distribution
of link traffic resulting from the reference models with that from the
empirical data.  The data reveals a very wide power law for
$P(\omega)$ with degree $1.9$.  This is consistent with our prior
results for host-level traffic~\cite{meiss08}.

The comparison with Page\-Rank and BookRank in Fig.~\ref{fig:traf_link} is a vivid
illustration of the diversity of links when we consider their
probability of actually being clicked.  A rough argument may again
help to make sense of the PageRank reference model's poor performance
at reproducing the data.  If we disregard teleportation, the traffic
to a page is roughly proportional to the page in-degree.  The traffic
expected on a link would be thus \emph{proportional} to the traffic to
the originating page and \emph{inversely proportional} to the
out-degree of the page if we assume that links are chosen uniformly at
random.  Since a node's in-degree and out-degree are equal in our
simulated graphs, this would lead to a link traffic that is
independent of the degree and therefore essentially constant for all
links.  This is reflected in the quickly decaying distribution of link
traffic for Page\-Rank.  
In the case of BookRank, the stronger heterogeneity in the probability 
of visiting pages is reflected in a heterogeneous choice of links, resulting 
in a broad distribution that fits the empirical data well as shown 
in Fig.~\ref{fig:traf_link}.


\begin{figure}
\centerline{\includegraphics[width=\columnwidth]{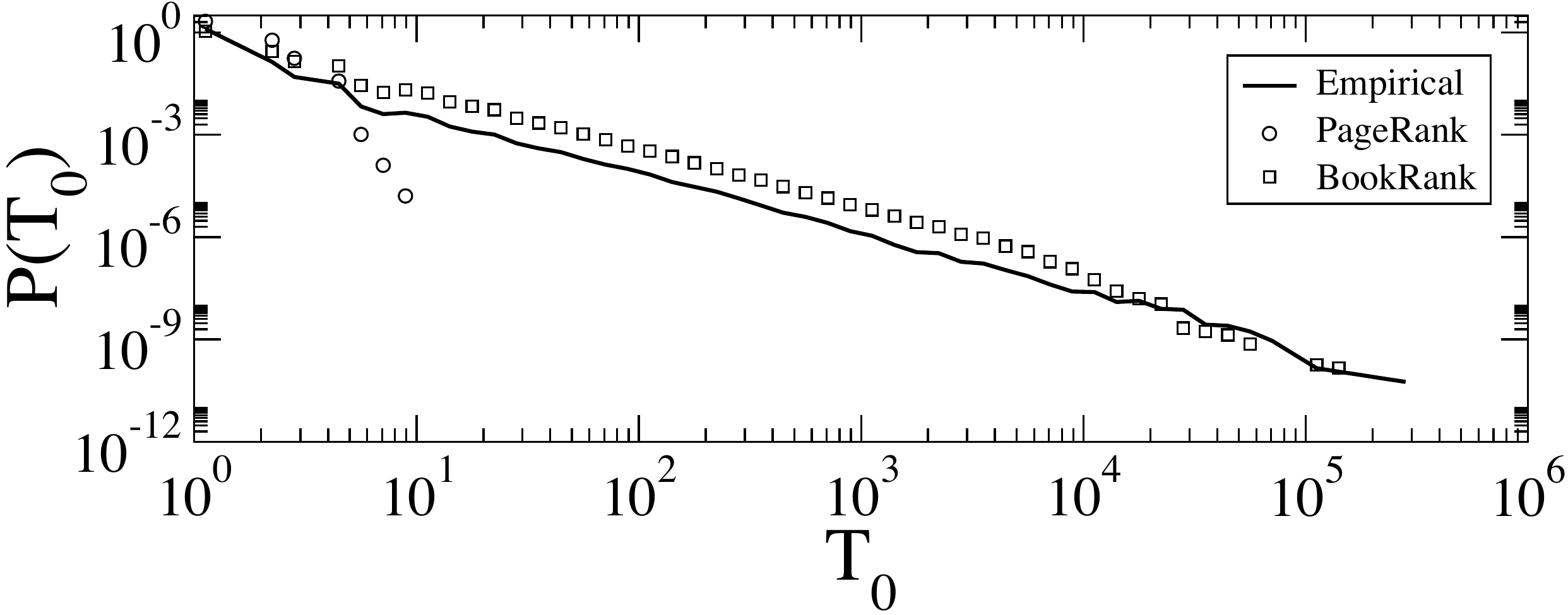}}
\caption{Empirical distribution of traffic originating from jumps (page requests with empty referrer) versus baselines.}
\label{traf_jump}
\end{figure}

Our empirical data in Fig.~\ref{traf_jump} show clearly that all pages
are not equally likely to be chosen as the starting point of a browsing
session.  Their popularity as starting points is roughly distributed
as a power law with an exponent close to 1.8 (consistent with prior
results for host-level traffic~\cite{meiss08}), implying a diverging
variance and mean when the number of sessions considered increases.
While not unexpected from a qualitative point of view, this
demonstrates how off the mark is one of the basic hypotheses
underlying the Page\-Rank class of browsing processes, namely uniform
teleportation.  Page\-Rank assumes a uniform probability for a
page to be chosen as a starting point, and its failure to reproduce
the empirical data is evident in Fig.~\ref{traf_jump}.  
The bookmarking mechanism, on the other hand, captures well the non-uniform 
probability of starting pages, so that the distribution generated by 
BookRank is a good match to the empirical data, as shown in Fig.~\ref{traf_jump}.


\begin{figure}
\centerline{\includegraphics[width=\columnwidth]{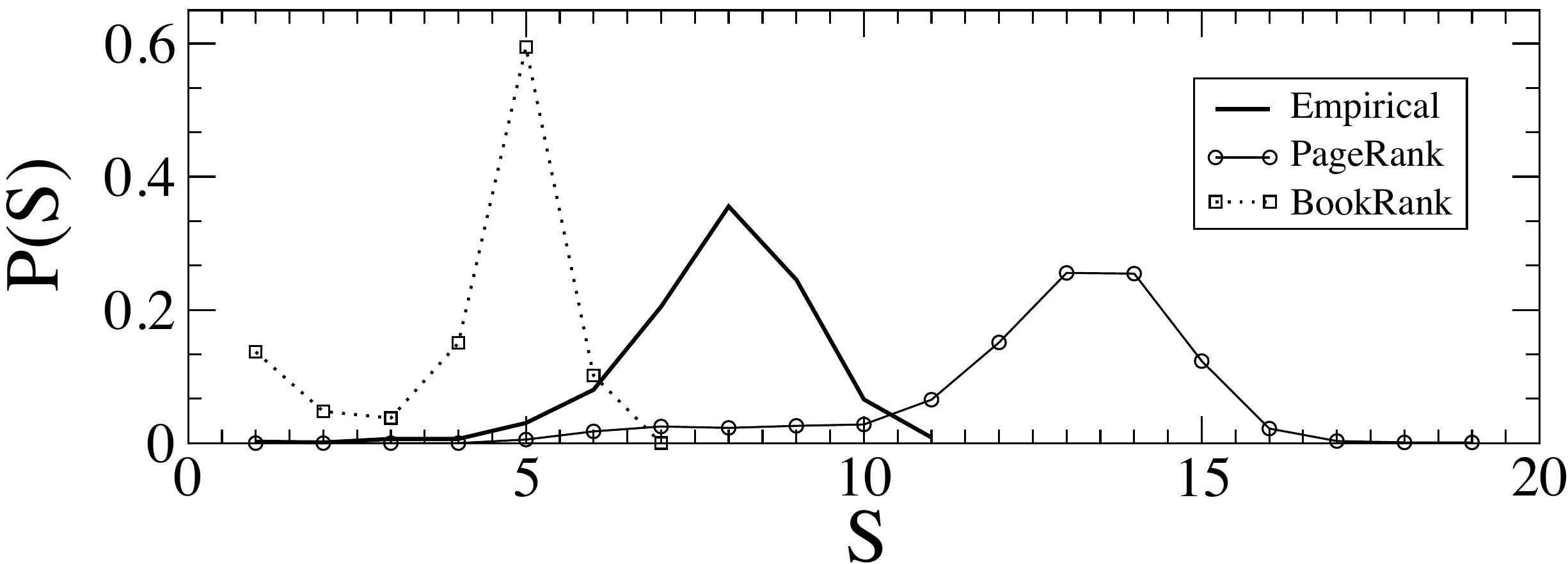}}
\caption{Empirical distribution of user entropy versus baselines.}
\label{fig:entropy}
\end{figure}

\begin{figure}[t]
\centerline{\includegraphics[width=\columnwidth]{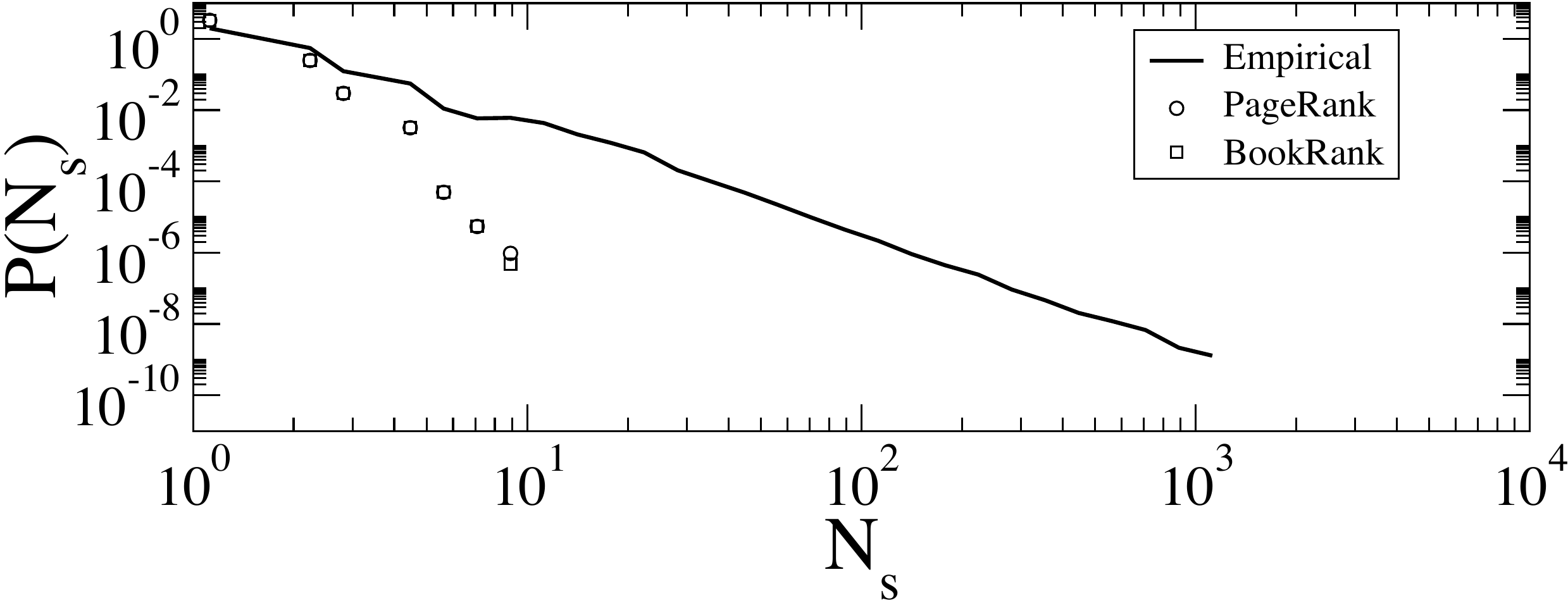}}
\caption{Empirical distribution of session size (unique pages per session) versus baselines.}
\label{size}
\end{figure}

\begin{figure}[t]
\centerline{\includegraphics[width=\columnwidth]{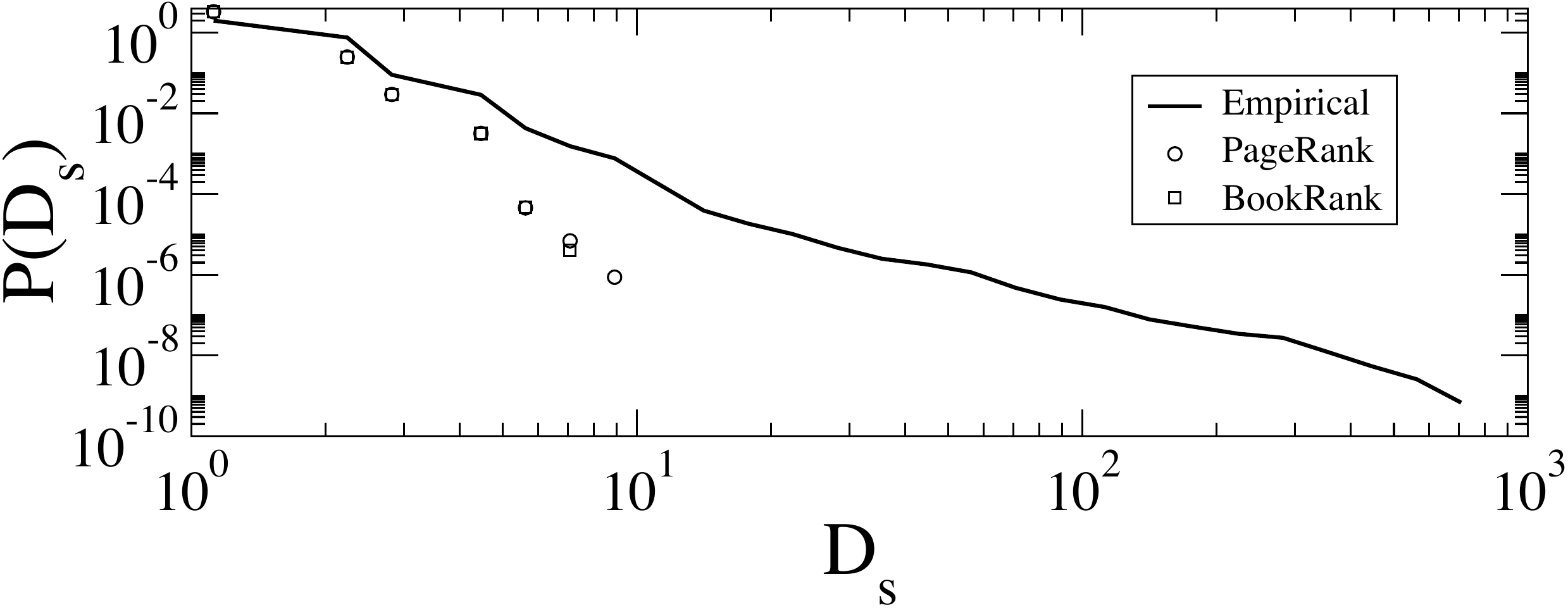}}
\caption{Empirical distribution of session depth versus baselines.}
\label{depth}
\end{figure}

We now turn from the aggregate properties of the system
and attempt to characterize individual users. 
The simplest hypothesis would be that the broad distributions 
characterizing aggregate user behavior are a reflection of
extreme variability within the traffic generated by single users, thus concluding 
that there is no such thing as a ``typical'' user from the point 
of view of traffic generated.  To capture how diverse is the behavior 
in a group of users, we adopt Shannon's information entropy of a user as defined above. 
Entropy directly measures the focus of a user's interests, offering 
a better probe into single user behavior than, 
for instance, the number of distinct pages visited; two users who have visited 
the same number of pages can have very different measures of entropy.
Given an arbitrary number of visits $N_v$, the entropy is maximum ($S
= N_v \log(N_v)$) when $N_v$ pages are visited once, and minimum
($S=0$) when all visits have been paid to a single page.  The
distribution of entropy across users is shown in
Fig.~\ref{fig:entropy}.  We observe that the reference Page\-Rank model
produces higher entropy than observed in the empirical data.  One can
interpret this by the way a Page\-Rank walker picks starting pages with uniform 
probability, while a real user is more likely to start from a previously visited
page, and therefore to revisit neighboring pages. BookRank is more similar to 
such repetitive behavior, and indeed we observe lower entropy values 
in Fig.~\ref{fig:entropy}. However, BookRank underestimates the entropy as 
well as its variability across users.  


\begin{figure*}
\centerline{
\includegraphics[width=0.44in]{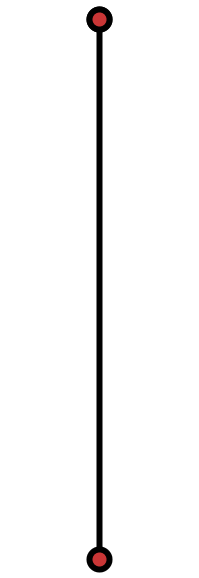}
\includegraphics[width=\onethird]{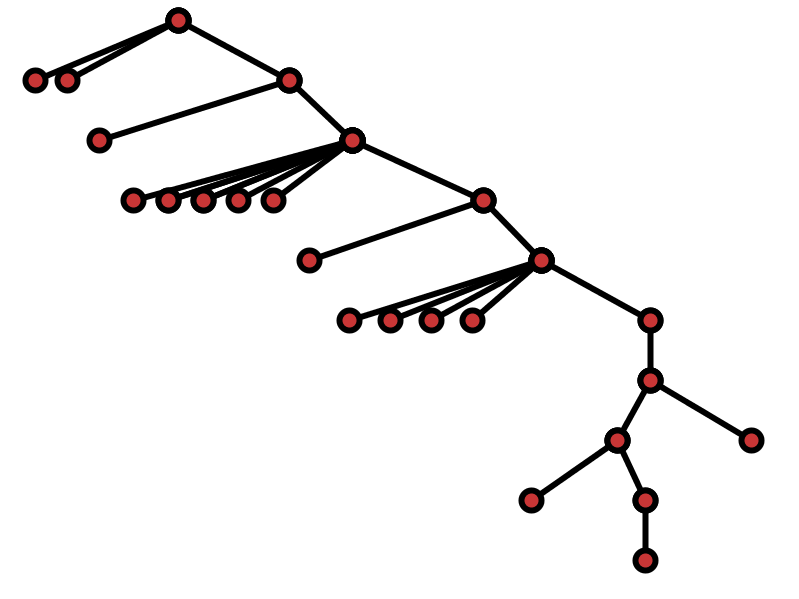}
\includegraphics[width=\onethird]{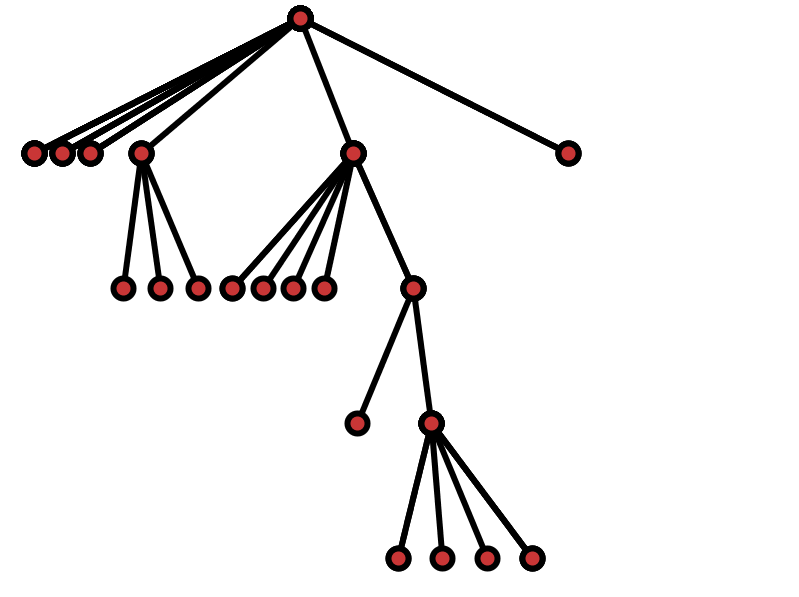}
\includegraphics[width=\onethird]{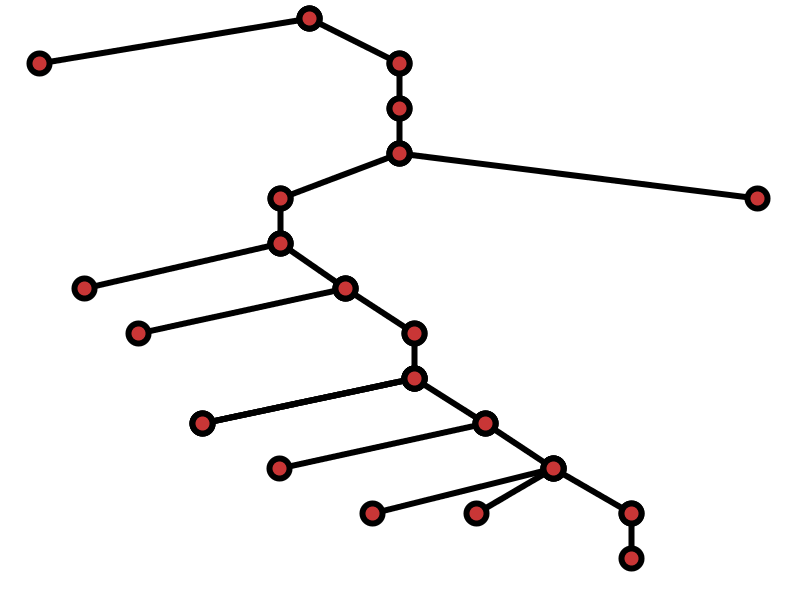}
}
\centerline{
\includegraphics[width=0.44in]{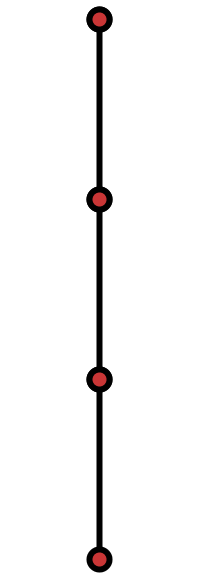}
\includegraphics[width=\onethird]{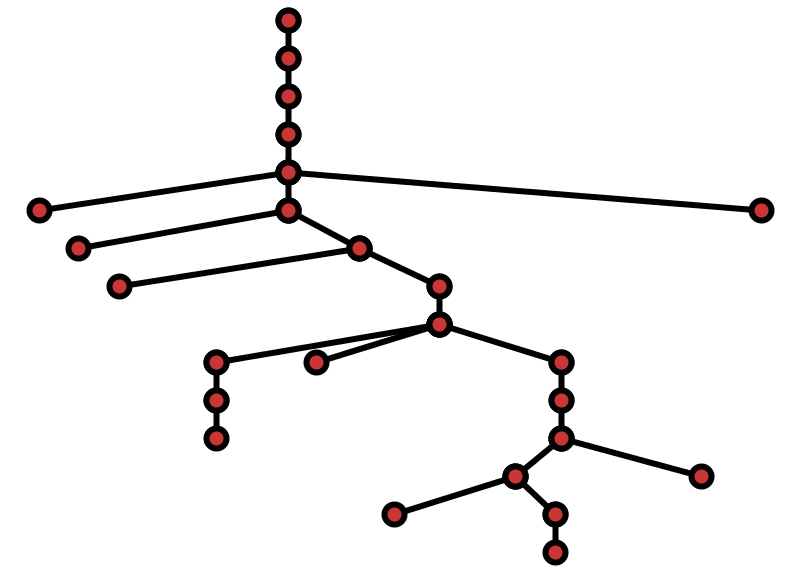}
\includegraphics[width=\onethird]{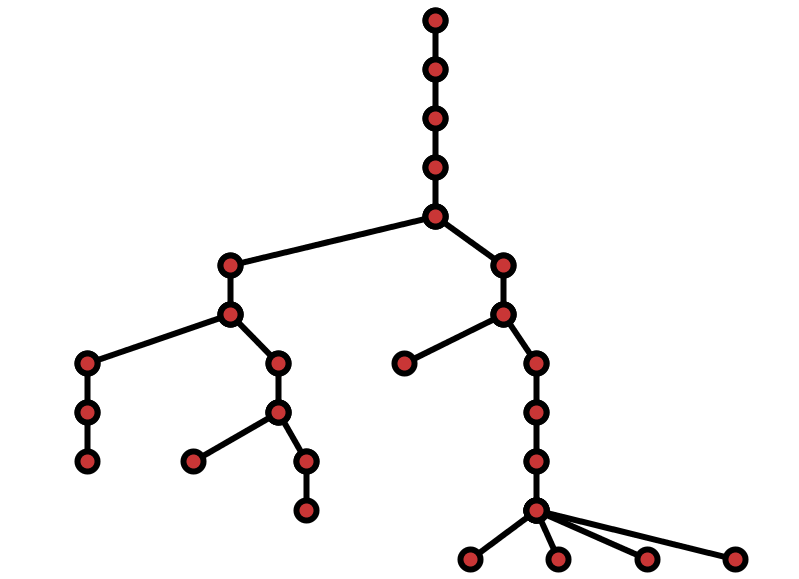}
\includegraphics[width=\onethird]{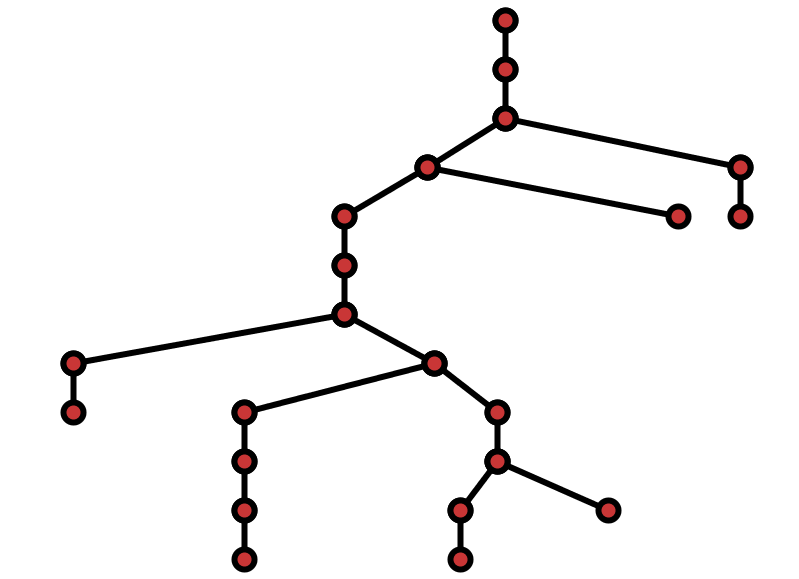}
}
\caption{Representation of a few typical and representative session trees from the empirical data (top) and from the ABC model (bottom). Animations are available at \protect\url{cnets.indiana.edu/groups/nan/webtraffic}.}
\label{cartoons}
\end{figure*}


Finally, we can consider the distributions that characterize logical
sessions, namely the size (number of unique pages) and depth (distance
from a session's starting page) distributions.  
Figs.~\ref{size}~and~\ref{depth} show that both empirical distributions are rather broad,
spanning three orders of magnitude, which is a surprisingly large
proportion of very long sessions.  In contrast, both Page\-Rank and BookRank
reference models generate very short sessions.
The probabilistic teleportation mechanism that determines when a
PageRank walker starts a new session is incapable of capturing broadly
distributed session sizes.  In fact, session size is upper-bounded by
the length $\ell$ (number of clicks) of a session, which exhibits a
narrow, exponential distribution $P(\ell) \sim (1-p_t)^{\ell}$.
Note that the exponentially short sessions are not inconsistent with the 
high entropy of Page\-Rank walkers (Fig.~\ref{fig:entropy}), which is a result 
of the frequent jumps to random targets rather than the browsing behavior.


\section{ABC model}
          
The empirical analysis in the previous section demonstrates that a 
more sophisticated model of user behavior is needed to capture 
individual navigation patterns. We build upon the BookRank model 
by adding two additional ingredients. 

First, we provide agents with a \emph{back button.}  
A backtracking mechanism is needed to capture the 
tree-like structure of sessions (see also top row of Fig.~\ref{cartoons}). 
Our data also indicates that the incoming and outgoing traffic of a
site are seldom equal.  Indeed, the ratio between incoming and
outgoing clicks is distributed over many orders of 
magnitude~\cite{meiss08}.  This violation of flow conservation cannot
be explained by teleportation alone, demonstrating that users'
browsing sessions have many branches.  Finally, our prior results show
that the average node-to-depth ratio of session trees is almost two. 
All of these observations are consistent with the use of tabs and
the back button.  Other studies have shown that the back button is
used frequently~\cite{Cockburn2001, Tauscher1997}. We therefore 
use the back button to model any branching behavior. 



The second ingredient has to do with the fact that the BookRank model fails to
predict individual statistics: all agents are identical, session size has 
a narrow, exponential distribution, and the comparison with the empirical 
entropy distribution is unsatisfactory.
In the real world, the duration of a session depends on the intentions
(goals) and interests of a user; different users have different interests.  
Visiting relevant pages, those whose topics match the user's interests, 
will lead to more clicks and thus longer sessions.  We therefore 
introduce the elements of different agents with distinct \emph{interests} 
and page \emph{topicality} into the model.  The idea is that an agent spends
some attention when navigating to a new page, and attention is gained
when visiting pages whose topics match the user's interests.  To model
this process, we imagine that each agent stores some ``energy'' (units
of attention) while browsing.  Visiting a new page incurs a higher
energy cost than going back to a previously visited page.  Known pages
yield no energy, while unseen pages may increase the energy store by
some random amount that depends on the page's relevance to the agent.
Agents continue to browse until they run out of energy, whereupon they
start a new session.

We call the resulting model \emph{ABC} for its main ingredients:
agents, bookmarks and clicks.  Clicks are driven by the topicality of
pages and agent interests, in a way that is in part inspired by the
\emph{InfoSpiders} algorithms for topical crawlers~\cite{Menczer97b,
Menczer00, Menczer02TOIT}.  InfoSpiders were designed to explore the
Web graph in an adaptive and intelligent fashion, driven by the
similarity between search topics and page content.  Better matches led
to more energy and more exploration of local link neighborhoods.
Irrelevant pages led to agents running out of energy and dying, so
that resources would be allocated to more promising neighborhoods.  In
ABC, this idea is used to model browsing behavior.

\begin{figure}[t]
\centerline{\includegraphics[width=\columnwidth]{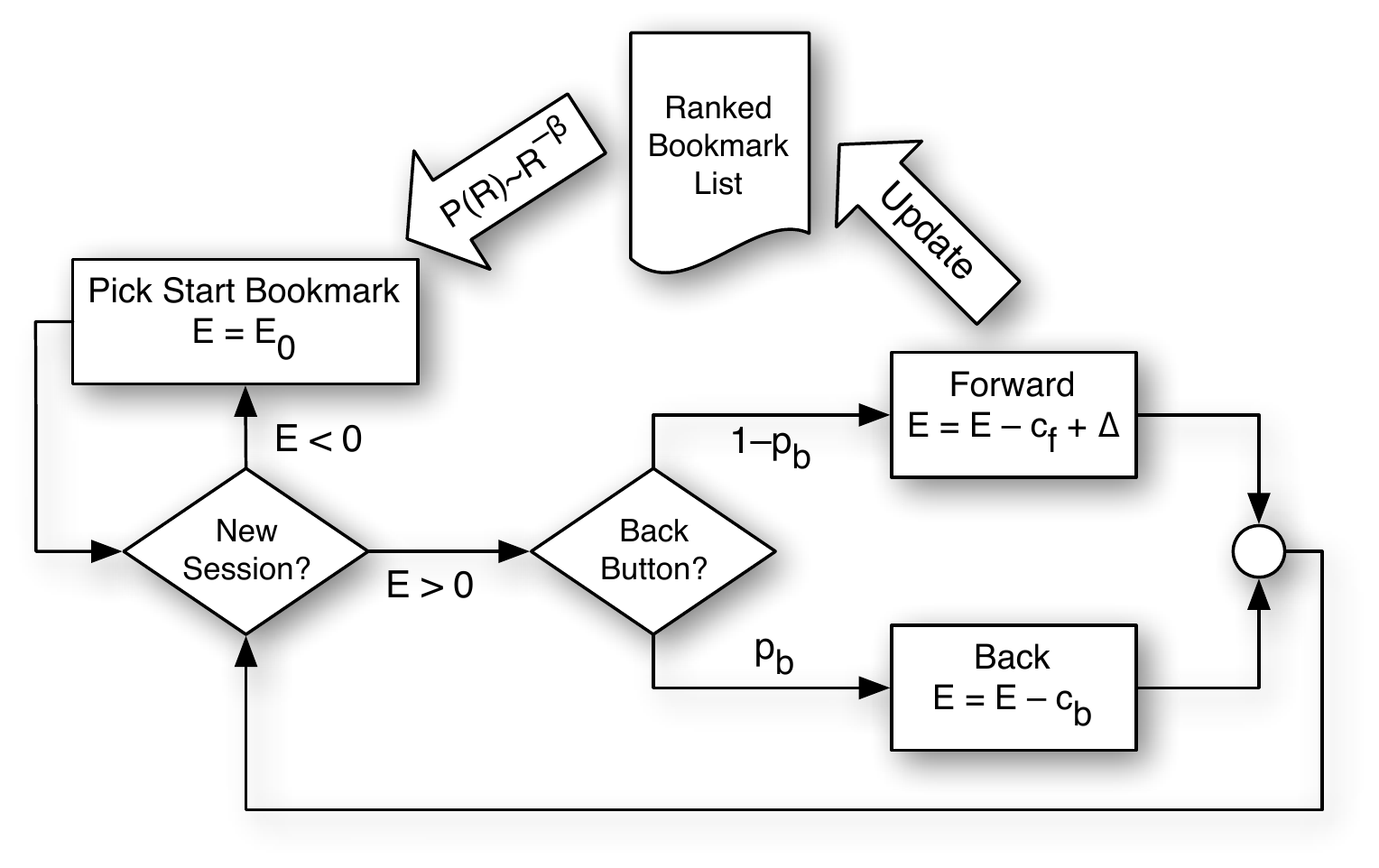}}
\caption{Schematic illustration of the ABC model.}
\label{models}
\end{figure}

The ABC model is illustrated in Fig.~\ref{models}.  Each agent starts
at a random page with an initial amount of energy $E_0$.  Then, for
each time step:

\begin{enumerate}

\item
If $E \leq 0$, the agent starts a new session by teleporting to a bookmark chosen as in BookRank.

\item
Otherwise, if $E > 0$, the user continues the current session,
following a link from the present node.  There are two alternatives: 

\begin{enumerate*}

\item With probability $p_{b}$, the back button is used, leading back to
the previous page.  The agent's energy is decreased by a fixed cost
$c_b$.

\item Otherwise, with probability $1-p_{b}$, a forward 
link is clicked with uniform probability. 
The agent's energy is updated to $E - c_f + \Delta$ where $c_f$ 
is a fixed cost and $\Delta$ is a stochastic value representing the 
new page's relevance to the user. 
As in BookRank, the bookmark list is updated with new pages and 
ranked by visit frequency.

\end{enumerate*}
\end{enumerate}

The dynamic variable $\Delta$ in the ABC model is a measure of relevance of a 
page to a user's interests.  The simplest way to model relevance is by a 
random variable, for example drawn from a Gaussian distribution.  In this case 
the amount of stored energy behaves as a random walk.  It has been shown that the session 
duration $\ell$ (number of clicks until the random walk reaches $E=0$) has a power-law tail 
$P(\ell) \sim \ell^{-\frac{3}{2}}$~\cite{Huberman98}. 
However, our empirical results suggest a larger exponent~\cite{meiss09}.  
More importantly, we know from empirical studies that the content
similarity between two Web pages is correlated with their distance in
the link graph, and so is the probability that a page is relevant with
respect to some given topic~\cite{Davison00, Menczer02maps,
Menczer01topo}.  Therefore, two neighbor pages are likely to be
related topically, and the relevance of a page $t$ to a user is
related to the relevance of a page $r$ that links to $t$.  To capture
such \emph{topical locality}, we introduce correlations between the
$\Delta$ values of consecutively visited pages.
For the starting page we use an initial value $\Delta_0 = 1$.  Then,
when a page $t$ is visited for the first time \emph{in a given
session,} $\Delta_t$ is determined by
\[
\Delta_t = \Delta_r (1 + \epsilon)
\]
where $r$ is the referrer page, $\epsilon$ is a random variable
uniformly distributed in $[-\eta, \eta]$ and $\eta$ is a parameter
controlling the degree of topical locality.  In a new session we
assume a page can again be interesting and thus provide the agent with
energy, even if it was visited in a previous session.  However, the
same page will yield different energy in different sessions, based on
changing user interests.

\section{Model evaluation} 

\subsection{Simulation of ABC model}

We ran two sets of simulations of the ABC model, in which agents navigate two 
distinct scale-free graphs. One (G1) is the artificial network discussed in 
\S~\ref{ref-models}.  Recall that $N = 10^7$ nodes and the degree distribution 
is a power law with exponent $\gamma = 2.1$ to match our data set. The second 
graph (G2) is derived from an independent, empirical, anonymous traffic data set.  
The data is obtained by extracting the largest strongly connected component from 
a traffic network generated by the entire Indiana University system population 
(about 100,000 people) \cite{meiss08}. This way there are no dangling links, 
but the nodes correspond to actual visited pages and the edges to actual 
traversed links. 
G2 is based on three weeks of traffic in November 2009; it has $N = 8.14 \times 10^6$ nodes and the same degree distribution with exponent $\gamma \approx 2.1$.

Within each session we simulate the browser's cache as discussed in \S~\ref{ref-models} so that 
we can measure the number of unique pages visited by the model agents and compare it with the empirical session size. 
%
%


The proposed models have various parameters. 
In prior work~\cite{fortunato06}, we have shown that the distribution 
of traffic with empty referrer generated by our models is related 
to the parameter $\beta$ (cf. BookRank description in \S~\ref{ref-models}). 
Namely, the distribution is well approximated by a power law 
$P(T_0) \sim T_0^{-\alpha}$, where $\alpha = 1 + 1/\beta$.  To match the 
empirical exponent $\alpha \approx 1.75$ we set the parameter 
$\beta = 1 / (\alpha - 1) = 1.33$. 
We also fit the back button probability $p_b = 0.5$ from the data.

The ABC model contains a few additional free parameters: the initial
energy $E_0$, the forward and backward costs $c_f$ and $c_b$, and the
topical locality parameter $\eta$.  The initial energy and the costs
are closely related, and together they control session durations.
%
%
We therefore set $E_0 = 0.5$ arbitrarily and use an energy balance
argument to find suitable values of the costs.  Empirically, the
average session size is close to two pages.  The net loss per click of an
agent is $-\delta E = p_b c_b + (1-p_b) (c_f - \langle \Delta
\rangle)$ where $\langle \Delta \rangle = 1$ is the expected value of
the energy from a new page.  By setting $c_f = 1$ and $c_b = 0.5$, we
obtain an expected session size $1 - (1-p_b) E_0 / \delta E = 2$
(counting the initial page).  In general, higher costs lead to shorter
sessions and lower entropy.
%
We ran a number of simulations to explore the sensitivity of the model 
to the parameter $\eta$, settling on 
$\eta=0.15$. 
Smaller values mean that all pages have similar relevance, and the
session size and depth distributions become too narrow.  Larger values
imply more noise (absence of topical locality), and the session
distributions become too broad.
The results shown below refer to this combination of parameters.

The number of users in the simulation, and the number of sessions for each user, are taken from the empirical data. 
Because the model is computationally intensive, we partitioned the simulated users into work queues of roughly equal session counts, which we executed in parallel on a high-performance computing cluster.

\subsection{Comparison of model with empirical data} 

The simulations of the ABC model users generate session trees that can
be compared visually to those in the empirical data, as shown in
Fig.~\ref{cartoons}.  For a more quantitative evaluation of our model,
we compare its results with empirical findings 
described in \S~\ref{study}.  For each of the distributions discussed
earlier, we also compare ABC with the reference BookRank model.
The latter is simulated on the artificial G1 network.


\begin{figure}
\centerline{\includegraphics[width=\columnwidth]{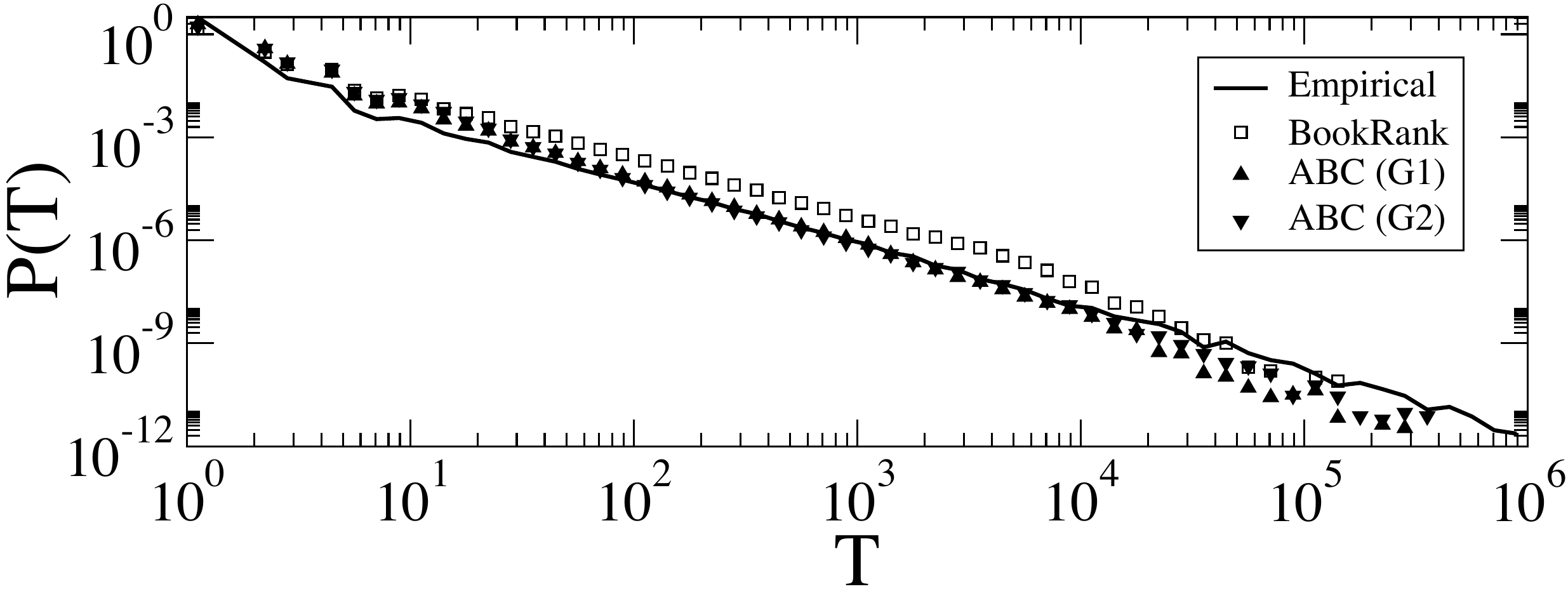}}
\caption{Distribution of page traffic generated by ABC model versus data and baseline.}
\label{traf_pages}
\end{figure}

\begin{figure}
\centerline{\includegraphics[width=\columnwidth]{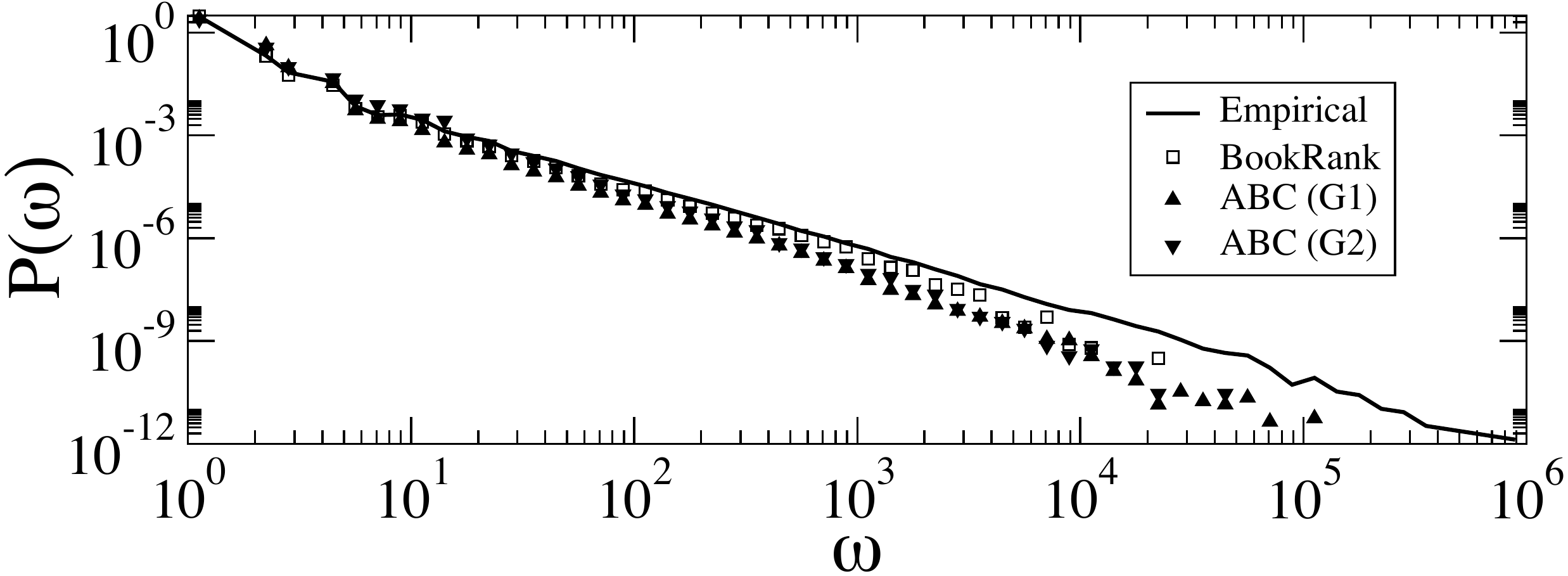}}
\caption{Distribution of link traffic generated by ABC model versus data and baseline.}
\label{traf_link}
\end{figure}

\begin{figure}[t]
\centerline{\includegraphics[width=\columnwidth]{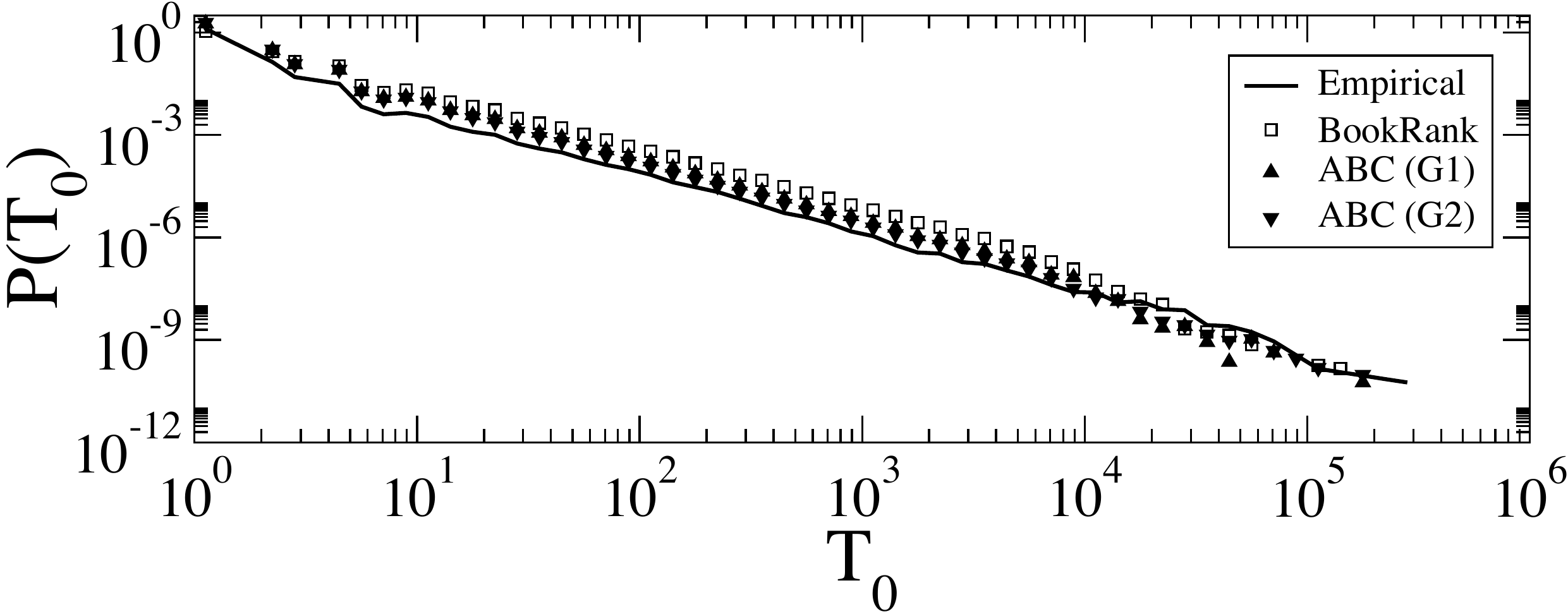}}
\caption{Distribution of traffic originating from jumps (page requests with empty referrer) generated by ABC model versus data and baseline.}
\label{traf_jump_model}
\end{figure}

A first aspect to check is whether the model is able to reproduce the
general features of the traffic distributions.  In
Fig.~\ref{traf_pages} we plot the number of visits received by each
page.  Agreement between the ABC model and data is as good as or better
than for the BookRank reference model.
%
%
Similarly, 
the distributions of link traffic (Fig.~\ref{traf_link}) and 
teleportation traffic (Fig.~\ref{traf_jump_model}) show that 
the ABC model reproduces the empirical data as accurately as BookRank. 

\begin{figure}
\centerline{\includegraphics[width=\columnwidth]{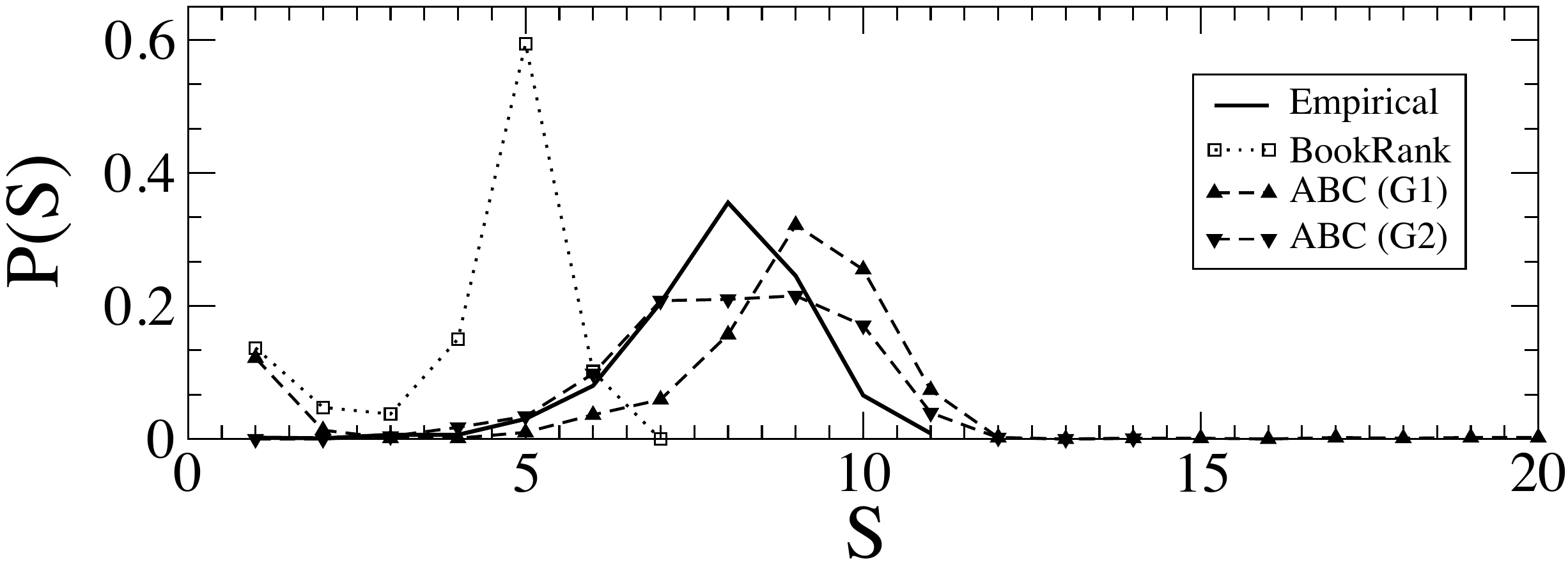}}
\caption{Distribution of user entropy generated by ABC model versus data and baseline. 
}
\label{entropy_model}
\end{figure}

%
The good agreement between both BookRank and ABC models and the data provides further
support for our hypothesis that the rank-based bookmark choice is a
sound cognitive mechanism to capture session behavior in Web browsing.

Let us now consider how our model captures the behavior of single
users.  The entropy distribution across users is shown in
Fig.~\ref{entropy_model}, where the model predictions are compared
with the distribution found in the empirical data.  
The ABC model yields entropy distributions that are 
somewhat sensitive to the underlying network, but that in any case fit 
the empirical entropy data much better than BookRank, in terms of both the 
location of the peak and the variability across users. 
This result suggests that bookmark memory, back button, and topicality are
crucial ingredients in explaining the focused habits of real users.
%


\begin{figure}[t]
\centerline{\includegraphics[width=\columnwidth]{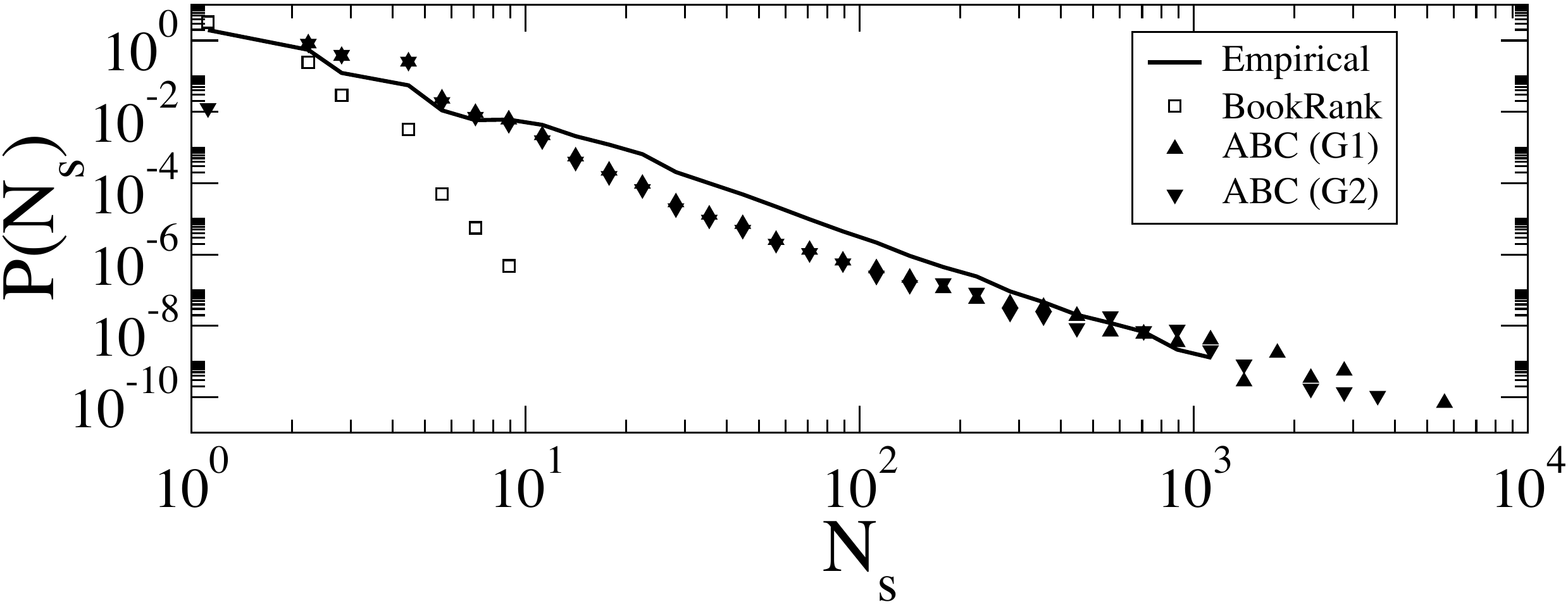}}
\caption{Distribution of session size (unique pages per session) generated by ABC model versus data and baseline. The distribution generated by the BookRank model cannot be distinguished from that of the PageRank baseline because they both use an identical exponential model of sessions.}
\label{size_model}
\end{figure}

Having characterized traffic patterns from aggregating across user
sessions, we can study the sessions one by one and analyze their
statistical properties.  In Fig.~\ref{size_model}, we show the
distribution of session size as generated by the ABC model.  
The user interests and topical locality
ingredients account for the broad distribution of
session size, capturing that of the empirical data much better that
the short sessions generated by the BookRank reference model.  Agents
visiting relevant pages tend to keep browsing, and relevant pages tend
to lead to other interesting pages, explaining the longer sessions.
We argue that the diversity apparent in the aggregate measures of traffic
is a consequence of this diversity of individual interests rather than
the behavior of extremely eclectic users who visit a wide variety of
Web sites --- as shown by the narrow distribution of entropy.


\begin{figure}[t]
\centerline{\includegraphics[width=\columnwidth]{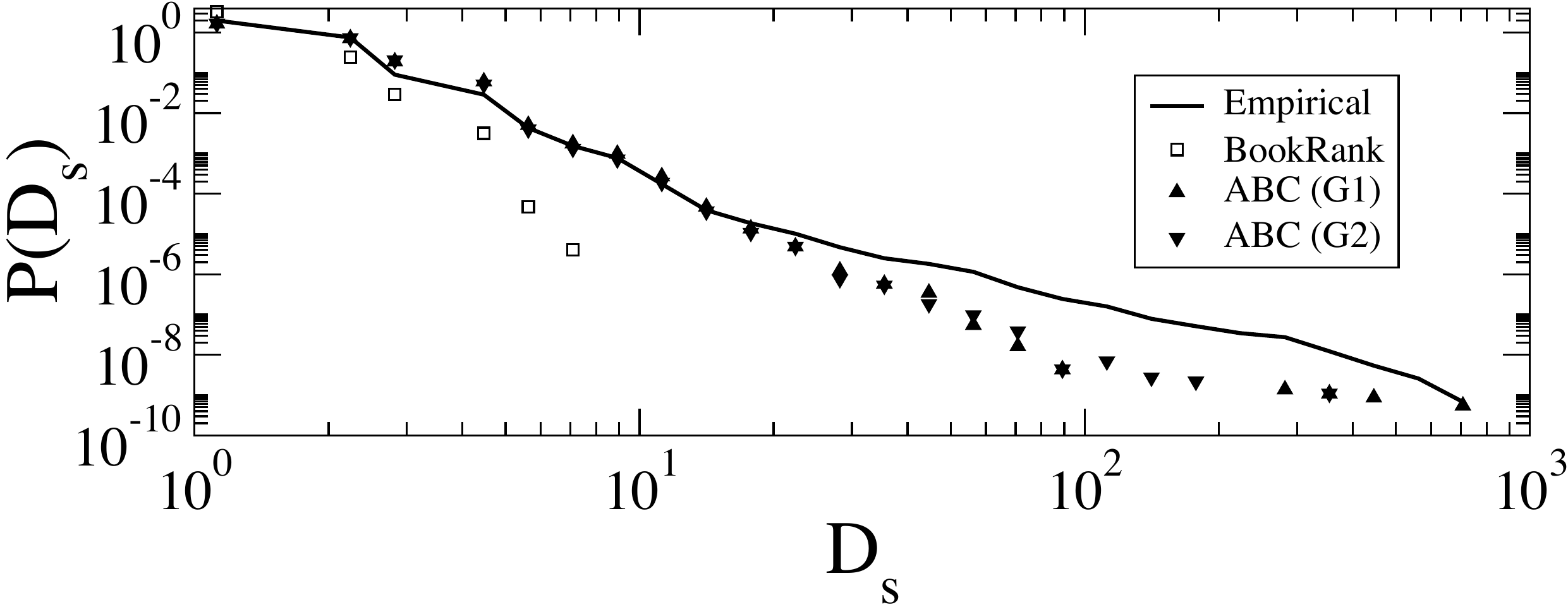}}
\caption{Distribution of session depth generated by ABC model versus data and baseline.}
\label{depth_model}
\end{figure}

The entropy distribution discussed above depends not only on session
length, but also on how far each user navigates away from the initial
bookmark where a session is initiated.  One way of analyzing this is
by the distribution of session depth, as shown in
Fig.~\ref{depth_model}.  The agreement between the empirical data and
the ABC model is excellent and significantly better than the one
observed with the BookRank baseline.  Once again topicality is shown
to be a key ingredient to understand real user behavior on the Web.



\section{Conclusions}

Several previous studies have shown that memoryless Markovian processes, such as 
PageRank, cannot explain many patterns observed in real Web browsing.
In particular, the diversity of session starting points, the global 
diversity on link traffic, and the heterogeneity of session sizes. 
The picture is further complicated by the fact that, despite such 
diverse aggregate measurements, individual behaviors are quite focused. 
These observations call for a non-Markovian agent-based model that can help explain the 
empirical data by taking into account several realistic browsing behaviors. Here 
we proposed three novel ingredients for such a model. First, agents maintain 
individual lists of bookmarks (a memory mechanism) that are used as 
teleportation targets. Second, agents have access to a back button (a 
branching mechanism) that also allows to reproduce tabbed browsing
behavior. Finally, agents have topical interest that are matched by 
page content, modulating the probability that an agent continues to browse 
or starts a new session and thus allowing to capture heterogeneous session 
sizes. 

We have shown that the resulting ABC model is capable of reproducing 
with remarkable accuracy the aggregate traffic patterns we 
observe in our empirical data. More importantly, our model offers the  
first account of a mechanism that can generate key properties of logical 
sessions. This allows us to argue that the diversity apparent 
in page, link, and bookmark traffic is a consequence of the
diversity of individual interests rather than the behavior of very eclectic
users. Our model is able to capture, for the first time, the extreme heterogeneity 
of aggregate traffic measurements while explaining the narrowly focused 
browsing patterns of individual users. 

Of course, the ABC model is more complex than prior models such as PageRank or 
even BookRank. However, its greater predictive power suggests that 
bookmarks, tabbed browsing, and topicality are salient features in 
interpreting how we browse the Web. 
In addition to the descriptive and explanatory power of an agent-based model
with these ingredients, our results may lead the way to more sophisticated, 
realistic, and hence more effective ranking and crawling algorithms. 

The ABC model relies on several key parameters, and while we have attempted to make 
reasonable, realistic choices for some of these parameters and explored the 
sensitivity of our model with respect to some others, further work is needed 
to achieve a complete picture of the combined effect of the multiple 
parameters. We already know, for example, that some parameters such as 
network size, costs, and topical locality play a key role in modulating the 
balance between individual diversity (entropy) and session size.

While, in its current incarnation, the ABC model is a clear step in the right direction, it still shares
some of the limitations present in previous efforts.  The most notable example 
is the uniform choice among outgoing links from a page, which may be 
responsible for the imperfect match between the individual entropy values of our model agents 
and those of actual users.

Future work can also explore intrinsic, node-dependent jump probabilities to model
the varying intrinsic relevance that users attribute to sites; for 
example, well-known sites such as CNN or Wikipedia are likely to be seen as more reliable or credible
than unknown personal blogs.
Restrictions on the subset of nodes reachable by each user, in the form of disconnected components for individual sessions,
can be used to model different areas of interest.  

\section*{Acknowledgments}

The authors would like to thank the Advanced Network Management
Laboratory and the Center for Complex Networks and Systems Research,
both parts of the Pervasive Technology Institute at Indiana
University, and L.~J. Camp of the IU School of Informatics and
Computing, for support and infrastructure.  We also thank the network
engineers of Indiana University for their support in deploying and
managing the data collection system.  This work was produced in part
with support from the Institute for Information Infrastructure
Protection research program.  The I3P is managed by Dartmouth College
and supported under Award 2003-TK-TX-0003 from the U.S.~DHS, Science
and Technology Directorate.  BG was supported in part by grant
NIH-1R21DA024259 from the National Institutes of Health.  JJR is
funded by the project 233847-Dynanets of the European Union
Commission. This material is based upon work supported by the NSF
award 0705676.  This work was supported in part by a gift from Google.
Opinions, findings, conclusions, recommendations or points of view in
this document are those of the authors and do not necessarily
represent the official position of the U.S.~Department of Homeland
Security, Science and Technology Directorate, I3P, National Science
Foundation, Indiana University, Google, or Dartmouth College.

%
%
\bibliographystyle{plain}
\bibliography{www}  
%
%

\end{document}